\documentclass{elsart}
\usepackage{epsfig}
\usepackage{here}

\newcommand{\be}{\begin{equation}}
\newcommand{\ee}{\end{equation}}

\usepackage{amssymb}

\begin{document}

\begin {frontmatter}

\title{A low background facility inside the LVD detector at Gran Sasso}
\author[lngs]{F.Arneodo\corauthref{cor}},
\author[to]{W.Fulgione\corauthref{cor}}
\corauth[cor]{francesco.arneodo@lngs.infn.it, walter.fulgione@to.infn.it}
\address[lngs]{INFN - Laboratori Nazionali del Gran Sasso, Assergi, Italy}
\address[to]{Institute of Physics of Interplanetary Space, INAF, Torino, University of Torino and INFN-Torino, Italy}
\begin{abstract}
The Large Volume Detector (LVD) in the Gran Sasso Laboratory of INFN is an observatory mainly devoted to search for neutrinos from core collapse supernovae.
It consists of 1000 tons of liquid scintillator divided in 840 stainless steel tanks 1.5m$^3$ each.
In this letter  we present the possibility for LVD to work both as a passive shield and moderator for the low energy gamma and neutron background and as an active veto for muons and higher energy neutrons.\\
An inner region inside the LVD structure ("LVD Core Facility") can be identified, with a volume of about 30m$^3$, with the neutron background typical of an underground laboratory placed at a much deeper site. This region can be realized with a negligible impact on the LVD operation and sensitive mass.
The LVD Core Facility could be effectively exploited by a compact experiment for the search of rare events, such as double beta decay or dark matter. 
\end{abstract}

\end{frontmatter}

\begin{figure}[H]
    \begin{center}
      \vspace{0.cm}
      \includegraphics*[width=11.cm,angle=0,clip]{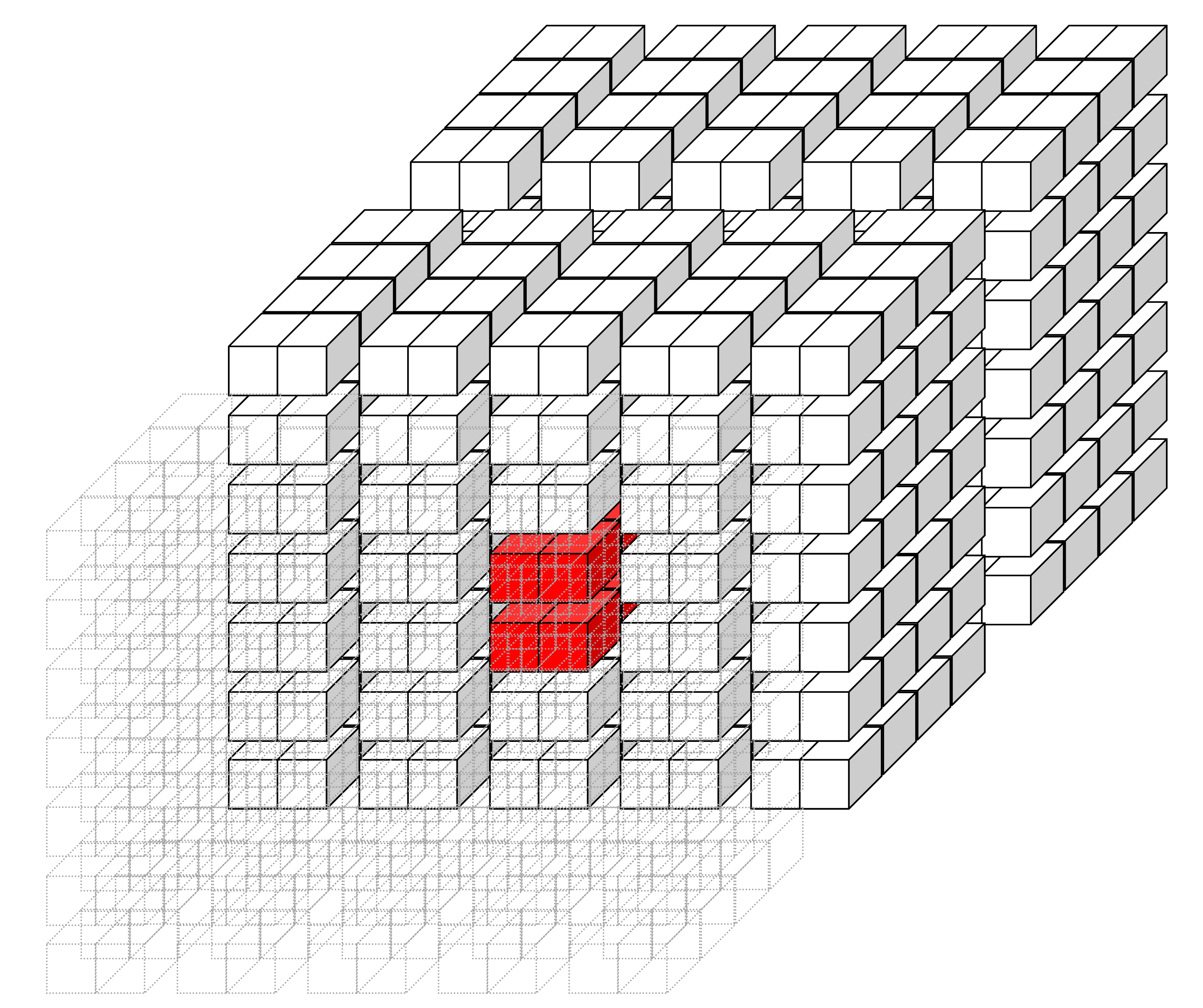}
    \end{center}
    \vspace{0.cm}
    \caption{{\bf A schematic view of the LVD detector, showing, in red, the modules corresponding to the LVD\_CF.}}
    \label{structure}
\end{figure}

\normalsize 
It is well known that the muon-induced fast neutron background limits the possibility of rare event searches like for neutrinoless double beta decay and WIMP dark matter.
Underground laboratories provide the overburden necessary to reduce this background, by attenuating cosmic-ray muons and their progenies.
If the depth of the underground laboratory is not enough to reach the necessary background reduction, the fast neutron flux 
can be shielded and/or actively vetoed.\\
In this letter we evaluate the shielding power of the LVD detector placed in the hall A of the INFN Gran Sasso National Laboratory \cite{LVD,on-line} at the equivalent vertical depth, relative to a flat overburden, of 3.1 km w.e..
We discuss the gamma and neutron background that would survive in a volume of 6 x 2 x 2.5 m$^3$, that we will call, from now on, the LVD Core Facility (LVD\_CF), see Fig.~\ref{structure}, 
placed inside the LVD detector which acts as an active veto and a passive shielding and moderator.\\

The gamma spectrum, up to E$_\mathrm{\gamma}$ = 3 MeV, has been measured inside and outside the LVD array.
The surviving $\gamma$ flux, in the LVD\_CF, should result attenuated of about a factor 20.
It is mainly due to the radioactive contamination of materials constituting LVD, with a possible contribution from the residual flux from the rock radioactivity.
To obtain an equivalent attenuation of the gamma flux a detector placed in hall A, outside LVD, would require about 5 cm of lead shielding.\\
LVD can act as a muon veto with respect to the LVD\_CF volume.
Assuming the  angular distribution from \cite{ang}, we expect a veto efficiency in excess of 99.8\% if a minimum energy release of 10 MeV in at least one tank is required and considering full efficiency at this energy.\\

Concerning the neutron background it can be divided into two energy regions:\\
- E$_\mathrm{n} < $ 10 MeV, arising from spontaneous fission mainly of $^\mathrm{238}$U and ($\alpha,n$) reactions due to radioactivity in the rock surrounding the LVD array and in the LVD components;\\
- $E_\mathrm{n} >$ 10 MeV, produced by  muon interactions in the rock or inside LVD.\\
Neutrons are detected in the LVD liquid scintillator by proton recoil. Because of  the quenching of the proton recoil light yield in scintillator, 
and taking into account the detector energy threshold (E$_\mathrm{th}$=4 MeV electron equivalent \cite{on-line}), the efficiency of LVD in tagging low energy neutrons ($E_n <$ 10 MeV) is quite low. 
Therefore,  in this energy range, LVD acts simply as a neutron moderator. 
Hydrocarbons are  very effective in suppressing the neutron flux; hence the empty space inside the LVD volume is the preferred way for neutrons to reach the LVD\_CF. 
However, such low energy neutrons  can be quickly absorbed using a polyethylene shielding of a reasonable thickness.\\
A completely different problem is represented by muon-induced high energy neutrons: they are the ultimate background limiting the sensitivity of an experiment searching for rare events in a deep underground laboratory \cite{Mei}\cite{Vitali}.\\ 
The muon interactions which produce these neutrons occur either inside the detector (see \cite{Menghetti}), or in the rock surrounding the experimental hall. 
These neutrons can be successfully eliminated as a source of background if  the parent muon,  its associated electro-magnetic cascades,  or the neutron itself, are detected by LVD.\\
Nevertheless, a small amount of muons  can go undetected  passing either  through a corridor or inside the iron structure. These "invisible" muons can generate neutrons that could reach the LVD\_CF without be vetoed.  Other high energy neutrons can be produced in the rock outside LVD and propagate through the horizontal corridor that cuts the LVD\_CF at its middle height.
In the following steps we will evaluate an upper limit to the neutron flux inside the LVD\_CF due to these two sources. 
\begin{enumerate}
\item We consider LVD as a parallelepiped with total volume, V$_\mathrm{tot}$ = 20x12x10 = 2400 m$^\mathrm{3}$ and total surface, S$_\mathrm{tot}$ = 1120 m$^\mathrm{2}$.\\
LVD is not uniform, as the corridors introduce gaps in the structure. Overall, the total volume is filled by: \\
- scintillator, V$_\mathrm{scint}$ = 1000/0.8 = 1250 m$^\mathrm{3}$;\\ 
- iron, V$_\mathrm{Fe}$ = 900/7.8 = 115 m$^\mathrm{3}$;\\
- air V$_\mathrm{air}$ = V$_\mathrm{tot}$-V$_\mathrm{scint}$-V$_\mathrm{Fe}\sim$ 1000 m$^\mathrm{3}$.\\
From the numbers above, it follows that, on the whole, the gaps between the scintillator volumes are filled $90\% $ by air and $10\%$ by iron.
Finally, the total external surface of the LVD corridors exposed to downward muons is $\sim17\%$ of the total LVD surface. Hence, we can reasonably argue that the percentage of "invisible" muons in LVD cannot be greater than 17\% of the total. This corresponds to an upper limit on the flux of "invisible" muons: 
$\Phi_\mathrm{invisible-\mu} < 0.17\cdot \Phi_\mathrm{\mu} < 4\cdot 10^{-9} cm^{-2}s^{-1}$.\\
These muons pass through the gaps between the scintillator volumes, i.e., they cross, roughly, air and iron in a 90/10 proportion.
From purely geometrical considerations, it follows that the  average track length of a muon  inside the LVD volume is $\bar L\sim$ 10 m, or, equivalently,  $\bar L~\leq 8 \cdot 10^{2}~ g~cm^{-2}$.
 Along this track, the  neutron production rate is \cite{Mei}: $Y_\mathrm{n} \sim 1.5 \cdot 10^{-3}~ n/(\mu~ g ~cm^{-2})$. This rate is dominated by Fe (the air contribution being $<$ 2\%). About 1/4 of the produced neutrons have energy greater than 10 MeV.\\
Only some of these neutrons will be able to reach the LVD\_CF.
Taking into account that inside LVD neutrons are attenuated with $\lambda_\mathrm{n} \leq $ 65 cm \cite{LVDn}, we consider only those neutrons produced inside the volume of thickness $\lambda_\mathrm{n}$ surrounding the LVD\_CF, that is 1/40 of the total LVD volume. \\
In this approximation the upper limit on the muon-induced neutron flux (E$_\mathrm{n}>$10MeV) produced by un-vetoed muons inside LVD and surviving at the LVD\_CF surface is:\\
$\Phi_\mathrm{n_\mathrm{~invisible-\mu}} (E_\mathrm{n}>10MeV) \leq \Phi_\mathrm{invisible-\mu} \cdot Y_\mathrm{n}/4 \cdot \bar L \cdot 1/40 = 3 \cdot 10^{-11} cm^{-2}s^{-1}$.\\
\item  To evaluate the contribution  of the neutrons propagating through the horizontal corridor we consider that:\\
- the surface of the LVD\_CF exposed to this corridor is: $\sim$10\% of its total surface;\\
- the mean solid angle corresponding to the horizontal corridor is $\Omega_{h}<1sr$.\\
It results a suppression factor with respect to the neutron flux emerging from the rock of the cavern
$> (0.1)^{-1} \cdot (\frac{\Omega_\mathrm{h}}{4\pi})^{-1} \sim 125$.
Assuming the muon-induced neutron flux, for E$_\mathrm{n}>$10 MeV, from \cite{Mei}, the flux surviving at the LVD\_CF will be:\\
$\Phi_\mathrm{n_\mathrm{~horizontal}} (E_\mathrm{n}>10MeV) \leq 7.3 \cdot 10^\mathrm{-10} / 125 = 5.8 \cdot 10^{-12} cm^{-2}s^{-1}$.\\
\end{enumerate}
The upper limit to the high energy, $E_\mathrm{n} >$ 10 MeV, neutron flux resulting from the sum of the neutrons emerging from the walls of the cavern and neutrons produced inside LVD by undetected muons, surviving to the LVD veto and reaching the LVD\_CF, is: 
\begin{equation}
\Phi_\mathrm{n_{~survived}}(E_\mathrm{n} > 10 MeV) < 3.6 \cdot 10^{-11} cm^{-2}~s^{-1}
\end{equation}
This estimate should be compared to the muon-induced neutron flux as predicted by \cite{Mei}, for Sudbury, at  the equivalent vertical depth of 6 km w.e.:  $\Phi_\mathrm{n_{~Sudbury}}(E_\mathrm{n} > 10 MeV) = 1.8 \cdot 10^{-11} cm^{-2}~s^{-1}$.\\

LVD has been designed with a highly modular structure. We can argue that moving some of its modules to create the Core Facility will have a negligible impact  on its live time. This is particularly important, not to spoil the  role of this detector  as a first-class observatory for stellar collapses. The removal of the modules from the Core Facility will, likewise, have a negligible impact on the total sensitive mass, as they constitute less than 2\% of the total.\\
The LVD Core Facility would have the adequate dimensions to host a next generation compact experiment for the search of rare events. 
The guest experiment would benefit of a low background environment and a large muon and neutron veto with a negligible cost in terms of dead time because the rate of singles, with E$\geq$10 MeV, in LVD is 0.03 Hz \cite{on-line}.
It will be exposed to  the background level of the deepest  sites in the world, while enjoying the ease of access of the Gran Sasso Laboratory.
On the other hand, the operation of the LVD detector will remain essentially untouched.\\
This would  create an unprecedented synergy between two underground  astro-particle experiments and their respective Collaborations.

\end{document}